# Non-destructive tomographic nanoscale imaging of ferroelectric domain walls

Jiali He[1], Manuel Zahn[1,2], Ivan N. Ushakov[1], Leonie Richarz[1], Ursula Ludacka[1], Erik D. Roede[1], Zewu Yan[3,4], Edith Bourret[4], István Kézsmárki[2], Gustau Catalan[1,5,6], Dennis Meier[1]

[1]Department of Materials Science and Engineering, Norwegian University of Science and Technology (NTNU), Trondheim, Norway.

[2]Experimental Physics V, Center for Electronic Correlation and Magnetism, Universität Augsburg, Augsburg, Germany.

[3]Department of Physics, ETH Zurich, Zurich, Switzerland.

[4]Materials Sciences Division, Lawrence Berkeley National Laboratory, Berkeley, USA.

[5]Catalan Institute of Nanoscience and Nanotechnology (ICN2), Campus Universitat Autonoma de Barcelona, Bellaterra, Catalonia.

[6]ICREA-Institucio Catalana de Recerca i Estudis Avançats, Barcelona, Catalonia.

**Extraordinary physical properties arise at polar interfaces in oxide materials, including the emergence of two-dimensional electron gases, sheet-superconductivity, and multiferroicity. A special type of polar interface are ferroelectric domain walls, where electronic reconstruction phenomena can be driven by bound charges. Great progress has been achieved in the characterization of such domain walls and, over the last decade, their potential for next-generation nanotechnology has become clear. Established tomography techniques, however, are either destructive or offer insufficient spatial resolution, creating a pressing demand for 3D imaging compatible with future fabrication processes. Here, we demonstrate non-destructive tomographic imaging of ferroelectric domain walls using secondary electrons. Utilizing conventional scanning electron microscopy (SEM), we reconstruct the position, orientation, and charge state of hidden domain walls at distances up to several hundreds of nanometers away from the surface. A mathematical model is derived that links the SEM intensity variations at the surface to the local domain wall properties, enabling non-destructive tomography with good noise tolerance on the timescale of seconds. Our SEM-based approach facilitates high-throughput screening of materials with functional domain walls and domain-wall-based devices, which is essential for monitoring during the production of device architectures and quality control in real-time.**

Ferroelectric domain walls are natural interfaces that separate regions with different polarization orientation. Because of their distinct local symmetry, electrostatics, and strain, the domain walls are a rich source for emergent electronic phenomena[1–3], including the formation of electronic inversion layers [4] and 2D electron gases [5]. The functional physical properties of the domain walls and their ultra-small feature size (down to sub-nanometer width) triggered the idea of developing domain-wall-based nanoelectronics, and different device concepts have been explored [6–9]. Initially, the walls received a lot of attention due to their spatial mobility, allowing to control electric currents by writing, repositioning, or erasing domain walls that act as reconfigurable interconnects [10]. More recently, spatially fixed domain walls moved into focus and it was shown that they can be used to emulate the behavior of electrical components, including digital switches [4] and AC-to-DC converters [11]. Thus, the domain walls themselves have turned into devices, facilitating innovative opportunities for next-generation nanotechnology.

It is established that the polarization configuration at the ferroelectric domain walls plays a key role for their functional properties [12–15]. Measuring the domain-wall orientation that determines the local charge state, however, remains a major challenge. This is because domain walls are often not perfectly flat; they can change their orientation within the bulk, form three-dimensional (3D) networks, or exhibit complex nanostructures. Furthermore, not all domain walls intersect with the surface and can be orientated parallel to it, which makes them hard to detect. A breakthrough was the advent of nonlinear optical methods that enabled imaging of ferroelectric domain walls in 3D [16–19]. Their application, however, is restricted to systems with specific optical properties and the spatial resolution is in the order of hundreds of nanometers, whereas domain-wall roughening and bending often occur on much smaller length scales [20–22]. Tomographic microscopy approaches offer higher resolution [23–26], but data acquisition times are rather long; most crucially, the established tomography methods for domain-wall imaging are destructive. Thus, they are incompatible with fabrication processes for future domain-wall devices, which will require the option of high-throughput sampling and a non-destructive way for the testing of materials and device architectures.

Here, we demonstrate that otherwise hidden ferroelectric domain walls in surface-near regions can be visualized and analyzed by scanning electron microscopy (SEM), accessing a depth of up to several hundreds of nanometers, as sketched in Figure 1a. Using the uniaxial ferroelectric $ErMnO_3$ as model system [27], we show that domain walls are detectable via characteristic SEM intensity variations, providing detailed information about the position and charge state of hidden walls. Based on surface and cross-sectional data, we derive a general model that relates the measured SEM contrast to the location and orientation of domain walls below the surface, allowing to reconstruct their structure with nanoscale spatial precision.

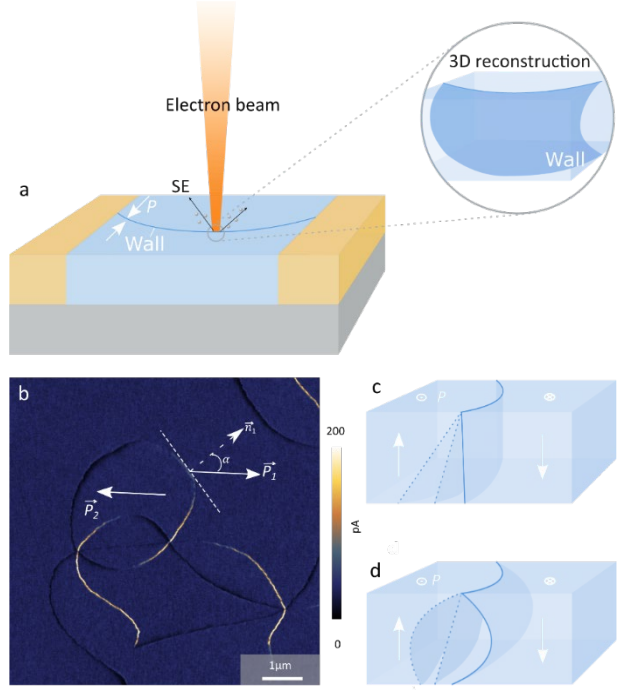

**Figure 1 | SEM tomography concept and domain wall structure of the model system ErMnO$_3$. a**, Secondary electrons (SE) carry rich information about the electronic material properties [28] at the surface and in near-surface regions. This sensitivity opens the door for SEM-tomography of ferroelectric domain walls, allowing to reconstruct their position, orientation, and charge state based on SEM intensity variations. **b**, cAFM image gained on the surface of our model system, ErMnO$_3$, with in-plane polarization $\vec{P}$ as indicated by the white arrows (acquired with a CDT-NCHR-10 probe tip at a bias voltage of 3 V applied to the back electrode). Tail-to-tail domain walls exhibit enhanced conductance relative to the bulk (bright), whereas reduced conductance is observed at head-to-head domain walls (black). The local domain wall charge state can be estimated based on Equation (1) by measuring the angle α between the wall normal $\vec{n_1}$ and the direction of $\vec{P_1}$. **c,d**, Illustrations showing domain walls in the near-surface region. Domain walls can exhibit different inclination angles (**c**) or pronounced curvature effects (**d**), which is not visible from surface-sensitive measurements alone.

ErMnO$_3$ naturally forms a 3D network with neutral (side-by-side), positively (head-to-head) and negatively (tail-to tail) charged domain walls [15]. The domain walls have been studied intensively and their fundamental physical properties are well understood [4, 11, 15, 23, 29–34], which makes the material an ideal model system for this work. At the tail-to-tail domain walls, mobile holes accumulate to screen the bound charges, $\rho_b$, and give rise to enhanced conductance as shown in Figure 1. Figure 1b presents a conductive atomic force microscopy (cAFM) image, where tail-to-tail walls appear as bright lines, indicating an about four times higher conductance than the ±$P$ domains they separate. In contrast to the tail-to-tail walls, reduced conductance is observed at head-to-head domain walls (black lines in Figure 1b), owing to a depletion of hole carriers as explained in detail in ref. [15]. The local charge state of the domain walls can be estimated based on their orientation relative to the polarization $\vec{P}$ of the adjacent domains

$$\rho_b = (\vec{P_1} - \vec{P_2}) \cdot \vec{n_1} = 2P \cdot \cos\alpha , \qquad (1)$$

with $\vec{P_1} = P$ in domain 1, and $\vec{P_2} = -P$ in domain 2 (the domain wall normal unit vector $\vec{n_1}$ points from domain 2 to domain 1). $\alpha$ is the angle between the local wall normal $\vec{n}$ and $\vec{P_1}$, as illustrated in Figure 1b. In this approximation, however, the sub-surface structure of the domain wall is neglected, which can lead to substantial deviations between the calculated bound charge and the actual charge density. For example, it was observed that nominally neutral domain walls (i.e., $\alpha = 90°$) in ErMnO$_3$ [29] and PbZr$_{0.2}$Ti$_{0.8}$O$_3$ [35] can exhibit enhanced or even metallic conductance, which was attributed to a non-zero inclination angle relative to the surface (Figure 1c). By performing cross-sectional experiments on LiNbO$_3$ [36], the impact of the inclination angle on the domain wall conductance was demonstrated, revealing that 10-15° tilting leads to a substantial enhancement. In addition, the domain wall curvature (Figure 1d) plays an important role as shown by focused ion beam (FIB) based 3D studies on ErMnO$_3$ [23]. In summary, these studies highlight the importance of the sub-surface structure of ferroelectric domain walls and the need for adequate characterization methods.

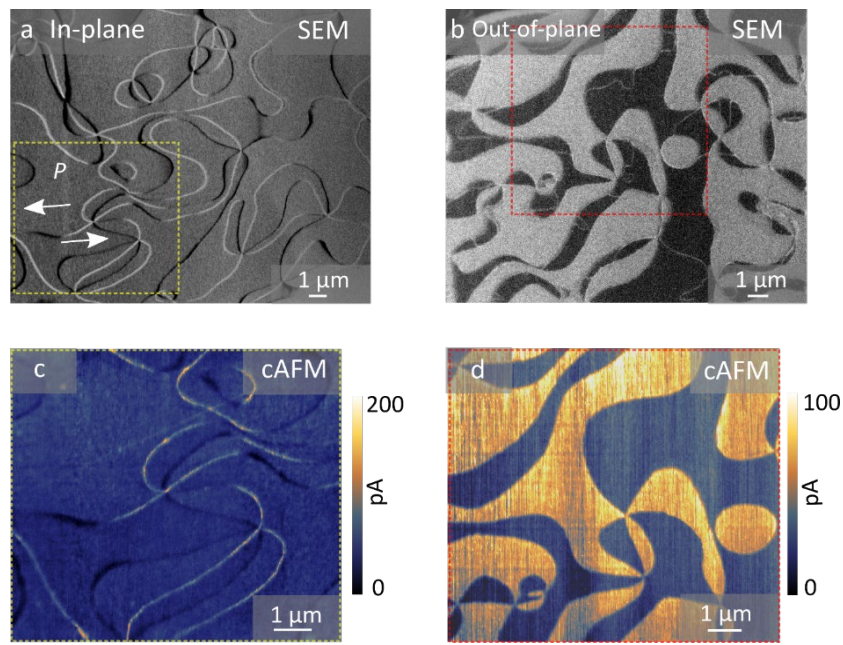

**Figure 2 | Correlated SEM and cAFM measurements on FIB-cut ErMnO$_3$ lamellas. a,** SEM image (2.0 kV, 0.4 nA, TLD) of a lamella with in-plane polarization (thickness ≈ 1 µm). Ferroelectric domain walls are visible as bright and dark lines. **b,** SEM image (2.0 kV, 0.1 nA, TLD) of a lamella with out-of-plane polarization (thickness ≈ 1 µm), showing pronounced domain contrast. **c,** cAFM image recorded in the region marked in **a** (yellow dashed rectangle). **d,** cAFM image of the region marked by the red dashed rectangle in **b**. The cAFM images in **c** and **d** are recorded with a doped diamond tip (HA_HR_DCP) and a bias voltage of 22.5 V applied to the back electrode.

An imaging technique that offers great potential for domain wall research in ferroelectrics is scanning electron microscopy (SEM). SEM has widely been applied for imaging domains and domain walls of ferroelectric materials, including BaTiO$_3$ [37], Gd$_2$(MoO$_4$)$_3$ [38], LiNbO$_3$ [39], and $R$MnO$_3$ ($R$ = Y, Er) [40, 41]. Although SEM is usually considered a surface-sensitive technique on account of the shallow escape depth of secondary electrons [42], it is also known that near-surface regions can play a crucial role for the emergent SEM contrast [43, 44]. The latter provides an as-yet-unexplored opportunity for minimally invasive analysis of the near-surface nanostructure of ferroelectric domain walls and their electronic properties.

To explore this possibility, we perform correlated SEM and cAFM measurements on lamellas which we extracted from an ErMnO$_3$ single crystal [45] with a focused ion beam (FIB), applying the same procedure as outlined in ref. [46]. Figure 2a and b show representative SEM images gained with the through-lens detector (TLD) detector on lamellas with in-plane and out-of-plane polarization, respectively. Both lamellas have a thickness of about 1 µm and are mounted on a flat Si-wafer with 100 nm Au coating. For the sample with in-plane polarization (Figure 2a), domain walls are visible as bright and dark lines. The walls form characteristic six-fold meeting points, corresponding to structural vortex/anti–vortex pairs as explained elsewhere [31, 47]. A conductive atomic force microscopy (cAFM) image from the region marked by the yellow dashed rectangle in Figure 2a is displayed in Figure 2c, showing the same domain wall pattern as the SEM image. Based on the cAFM data, we can identify the bright and dark lines in Figure 2a as conducting tail-to-tail and insulating head-to-head domain walls, respectively. Going beyond previous studies – which achieved contrast in FIB-cut lamellas only in the high-voltage regime where all walls are conducting [46] – we here access the low-voltage regime, where only the tail-to-tail walls exhibit enhanced conductance [4]. The latter is an important step, because it demonstrates that domain walls in lamellas and single crystals exhibit the same behavior, i.e., the applied nanostructuring by FIB does not alter the electronic properties of the domain walls. Figure 2b and d present analogous measurements for the lamella with out-of-plane polarization. Based on the comparison of the cAFM and SEM data, we find that the more conducting -*P* domains are brighter than the insulating +*P* domains in SEM (see, e.g., refs. [11, 48] for details on the polarization-dependent transport behavior at the level of the domains). The data in Figure 2 allows for calibrating our SEM measurement, showing that (under the applied imaging conditions) bright/dark SEM contrast indicates enhanced/reduced conductance. Note that this calibration step is crucial as the domain wall contrast in SEM depends on the imaging parameters and can, e.g., invert depending on the acceleration voltage [49].

On a closer inspection of the SEM data in Figure 2a, we observe gradual changes in intensity on one side for several of the domain walls. This behavior is presented in Figure 3a, showing a head-to-head domain wall with an asymmetric intensity distribution in the adjacent domains as marked by the blue dashed line. Occasionally, gradual intensity variations also occur within the domains as seen in Figure 3b. Qualitatively the same features arise in SEM measurements on millimeter-thick single crystals (Figure 3c). Analogous to Figure 3a, several head-to-head domain walls exhibit a distinct contrast on one side (marked by the yellow dashed line in Figure 3c). Furthermore, we observe distinct intensity variations within one of the domains (red dashed line). These contrast variations cannot be explained based on the nominal charge state of the walls at the surface alone, indicating additional contributions.

To understand the origin of such additional contrast contributions, we use the FIB to cut a cross-section parallel to the white line in Figure 3c as illustrated in the inset to Figure 3d. Figure 3d presents the corresponding cross-sectional SEM image, where letters A and B indicate the same positions as in Figure 3c. Consistent with the SEM data gained on the surface, the cross-sectional measurement shows a conducting tail-to-tail wall (bright) that reaches the surfaces at point A and an insulating head-to-head wall (dark) that surfaces at point B. The head-to-head wall (DW2, yellow dots) has a surface inclination angle of about 25.2° and propagates in the direction in which the gradual contrast change is observed in Figure 3c. Furthermore, the cross-sectional image reveals an additional domain wall in the near-surface region (DW1), as well as several domain walls deeper in the bulk (i.e., ≳ 3 μm away from the surface) that run almost parallel to the surface until they merge in a vortex-like meeting point. Interestingly, we find that DW1 changes its charge state, going from insulating (dark) to conducting (bright), and the respective turning point coincides with the position where we observed the change in SEM intensity on the surface (see Figure 3c and d). These observations indicate a close relationship between the SEM contrast measured at the surface and the (hidden) charged domain walls in the near-surface region.

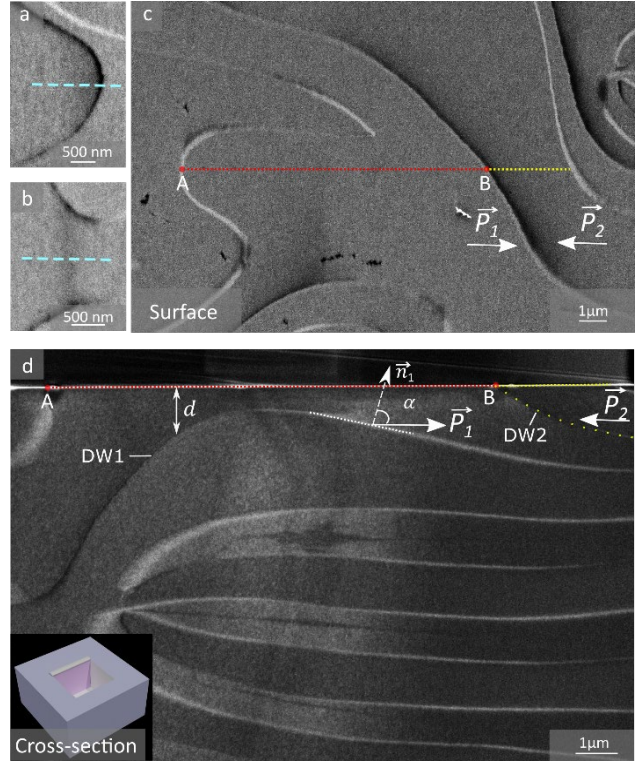

**Figure 3 | Correlated surface and cross-sectional SEM data. a,b,** Zoom-ins to the SEM image in Figure 2a, presenting examples of gradually varying SEM intensity in the vicinity of a domain wall (**a**) and within a domain (**b**). **c,** SEM data (1.5 kV, 0.1 nA, TLD) recorded on the surface of an ErMnO$_3$ single crystal. The image shows qualitatively similar features as in **a** and **b**. Along the red dashed line, a change in contrast is observed within the domain, whereas a gradual change in intensity on one side of the wall is measured in the region marked by the yellow dashed line. Labels A and B correspond to two positions where domain walls intersect with the surface, and white arrows show the polarization direction within the domains. **d,** Cross-sectional SEM image (2.0 kV, 0.1 nA, TLD) taken after FIB cutting a trench as sketched in the inset to **d**. Labels A and B mark the same positions as seen in c. Two domain walls in the near-surface region are highlighted (DW1 and DW2) and key parameters are presented ($d$ distance from the surface, $\vec{n_1}$ local normal to the domain wall, $\alpha$ angle between $\vec{n_1}$ and $\vec{P_1}$).

To relate the intensity measured at the surface to the position and structure of the domain walls in the near-surface region, we build a simple model. Based on the SEM data, within the first order Taylor expansion in $\rho_b$ and multipole-like expansion in $d$, we assume that variations in SEM intensity, $\Delta I$, scale with the density of bound charges ($\propto \rho_b$) and that related effects decrease with increasing distance between the wall and the surface ($\propto d^{-n}, n \in \mathbb{N}$), leading to

$$\Delta I \propto \frac{\rho_b}{d^n} \propto \frac{\cos \alpha}{d^n}, \tag{2}$$

($\cos \alpha > 0$ and $\cos \alpha < 0$ give tail-to-tail and head-to-head configurations, respectively). To derive the value of $n$, we extract the SEM intensity measured at the surface ($I_{\text{SEM}} = I_0 + \Delta I$) and the parameters $\alpha$ and $d$ from the SEM data in Figure 3c and d, respectively, considering two domain walls (DW1 and DW2) as explained in Supplementary Note 1 and Supplementary Fig. S1. This approach leads us to the conclusion that the experimentally observed dependence of $\Delta I$ on the distance between the wall and the surface is reproduced best for $n = 2$, i.e.,

$$\Delta I = \frac{A}{d^2} \cdot \cos \alpha \ (A = \text{const}). \tag{3}$$

A possible physical explanation for this relationship is the electrostatic effect of the domain wall bound charges on the secondary electrons [50]. As CASINO simulations [51] show, incident primary electrons (E = 1.5 keV at 0° tilt) lose 75% of their energy in the near-surface region with a depth of about 6.6 nm (maximum penetration depth ≲ 30 nm). The latter implies that the majority of secondary electrons is generated close to the surface, i.e., at a distance comparable to the parameter $d$ that describes the wall–surface distance in our model. One possible physical mechanism that leads to the domain-wall-related SEM contrast is electrostatic interaction. Considering secondary electrons and domain wall bound charges as points charges, $q_1$ and $q_2$, their interaction is described by Coulomb's law, $F \propto \frac{q_1 \cdot q_2}{d^2}$. When neglecting the impact of the free charge carriers and consider only the bound charge carriers at the domain wall, $q_2$, this translates into an electric field $E = -\frac{\partial F}{\partial q_1} \propto \frac{\rho_b}{d^2}$ from the bound charges that acts on the secondary electrons via electrostatic induction and, hence, influences the secondary electron yield [50]. It is important to note, however, that the SEM contrast formation is highly non-trivial in ferroelectrics with multiple possible contributions [28]; to clarify the microscopic origin, additional studies are required, which is beyond the scope of this work.

Importantly, our simple model reproduces the experimental data remarkably well (see Supplementary Fig. S2), corroborating that domain wall bound charges play a key role for the intensity distribution in SEM measurements. Most interestingly, the experiments demonstrate that Equation (3) holds for domain walls at distances up to ≈ 1.5 µm away from the surface, which is much larger than the penetration depth of the incident primary electrons. This finding reflects an outstanding sensitivity towards otherwise hidden domain walls and enables nanoscale 3D imaging of domain walls as we discuss in the following.

Figure 4a presents the SEM intensity (orange) measured at the surface above DW1. A zoom-in to the area of interest from which the line plot is generated is shown in the inset to Figure 4a. To derive a mathematical representation for $\Delta I$, we optimize a low order Taylor expansion of the domain wall shape $d(x)$, utilizing a basin-hopping optimization algorithm. This approach leads to the fit (back line) that is shown along with the SEM data in Figure 4a. The fit captures the main features seen in the SEM data and allows for calculating the domain-wall structure based on Equation 3. The result is displayed in Figure 4b, where the black curve represents the reconstructed domain wall. The orange curve corresponds to $d$-values extracted from the cross-sectional data (Figure 3d), which is shown for comparison to evaluate the quality of the reconstructed domain-wall structure. We find that based on the SEM map gained at the surface, we can determine the sign of higher derivatives $\mathrm{d}^{(n)}\, d/\mathrm{d}x^{(n)}$, which reveals whether the domain wall curvature is convex or concave in the near-surface region (the accuracy of the Taylor coefficients $a_n$ is about % $(n+1)^2 \cdot 3\%$, yielding good precision for the low-order coefficients, which are the most relevant ones for the reconstruction).

To go beyond the specific case of DW1, we next consider a hypothetical domain wall of arbitrary shape, corresponding to the profile (orange) shown in Figure 4d. Figure 4c presents the calculated SEM intensity (orange) with random Gaussian noise to emulate experimental fluctuations. Applying the same approach as for DW1, we fit the noisy intensity data, which leads to the black curve in Figure 4c. Based on this fit, we calculate the domain wall structure using Equation (3) (black reconstructed profile in Figure 4d). The reconstructed structure is in excellent agreement with the hypothetical domain wall of arbitrary shape that was used as input data, demonstrating the general validity of our reconstruction approach.

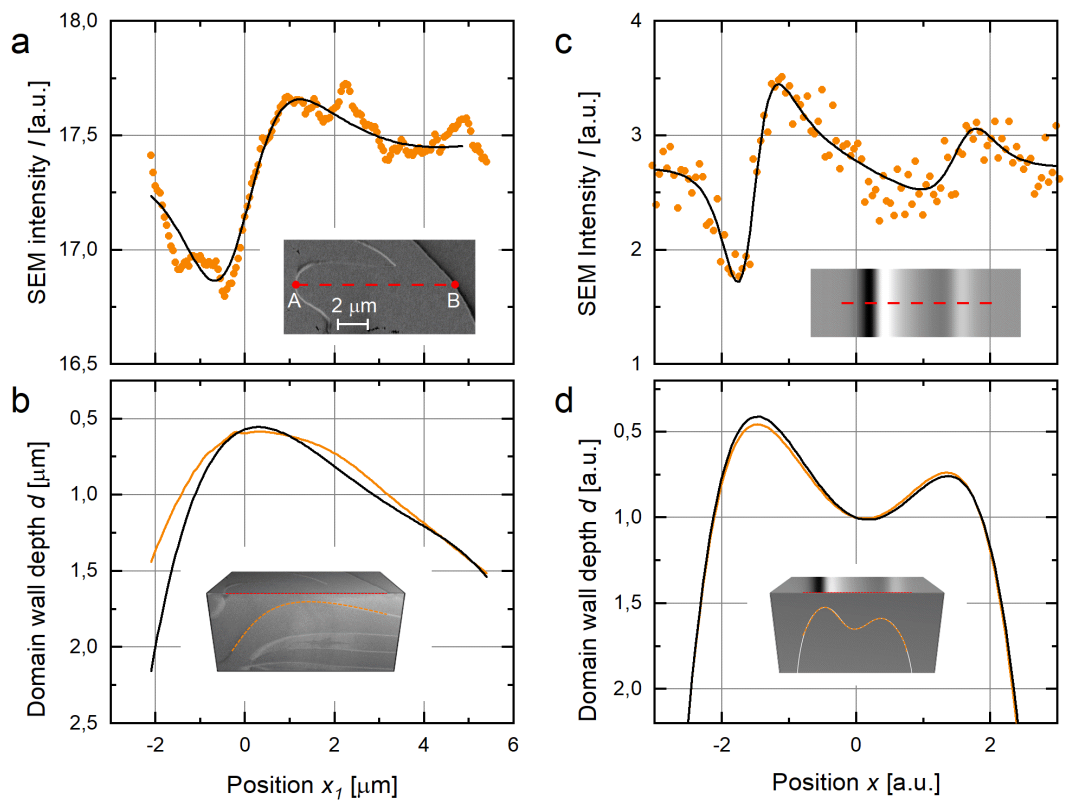

**Figure 4 | Reconstruction of the near-surface domain wall geometry from SEM intensity data. a,** SEM intensity recorded along the red dashed line (between A and B) shown in the inset and Figure 3c and calculated SEM intensity (black) based on the reconstructed domain wall in **b. b,** Shape of DW1 (see Figure 3d and inset) as measured from the cross section (orange) and reconstructed shape based on the SEM surface intensity (black). **c,** Simulated SEM intensity of an artificial domain wall of arbitrary shape overlayed with random noise for the reconstruction process to simulate random experimental fluctuations (orange) and SEM intensity from the reconstructed domain wall shape (black). **d,** Arbitrarily generated structure of the simulated domain wall (orange, input data) and its reconstructed shape (black).

Our results demonstrate the possibility to reconstruct the shape of ferroelectric domain walls in near-surface regions with nanoscale spatial resolution based on SEM maps. Key parameters of otherwise hidden domain walls, such as curvature and local charge state, thus become readily accessible within a single scan. In contrast to previously applied techniques with comparable resolution (e.g., 3D imaging by tomographic PFM or FIB), the SEM-based reconstruction process is non-destructive and much faster, allowing for image acquisition and data analysis on the time scale of seconds. On the one hand, this enables correlated experiments on individual domain walls or domain-wall networks with well-known orientation, curvature, and charge state, providing new opportunities for fundamental domain wall studies. Combined with machine learning and faster evaluation algorithms, dynamical electric-field, pressure- or temperature-driven changes in the 3D structure may be investigated in real time along with changes in the electronic domain wall response. On the other hand, SEM-tomography is of interest for domain-wall nanoelectronics, facilitating non-destructive high-throughput screening of materials with functional charged domain walls and domain-wall-based devices, which is essential for monitoring during the production of device architectures and quality control in real-time.

**Methods**

**Sample preparation.** High-quality ErMnO$_3$ single crystals were grown by using the pressurized floating-zone method [45]. The sample was then oriented by Laue diffraction, cut into 1 mm thick small pieces with polarization directions along one of the surface edges (in-plane polarization), and polished using silica slurry. Lamellae were extracted from the oriented single crystal using a Thermo Fisher Scientific G4UX Dual-beam FIB-SEM and subsequently mounted onto a flat Si wafer substrate coated with 100 nm of Au. Utilizing a pre-tilted stub with 45° tilting angle, the lamellas were polished using a Ga$^+$ ion beam at a glancing angle. The polishing process involved gradually decreasing the ion beam current to 90 pA. Subsequently, a 5kV Ga$^+$ ion beam was used to strip away a thin surface layer (about 30 nm), aiming to remove the ion beam damage layer. In addition to this established procedure as detailed in ref. [46], a final polishing step was introduced using an Argon ion beam polisher Gatan Precision Ion Polishing system (PIPS II). This procedure was conducted at liquid N$_2$ temperature, using beam voltage of 1 kV, 0.5 keV, and 0.3 keV, sequentially, at a beam operating angle of 6°.

**Scanning probe and scanning electron microscopy.** cAFM measurements were carried out using a Cypher ES environmental AFM (Oxford instruments) with diamond-coated AFM probe tips CDT-NCHR-10 and HA_HR_DCP as specified in the figure captions. SEM imaging was performed using the same Thermo Fisher Scientific G4UX Dual-beam FIB-SEM as for the lamella preparation / cross-sectioning. SEM images were captured by a TLD (Through the-lens detector) detector, using a beam acceleration voltage close to the charging equilibrium point. Specific beam parameters for each image can be found in the figure captions.

**Domain wall reconstruction.** To obtain the domain wall geometry from Figure 3d as shown in Figure 4b, a polygonal chain was initiated with a manually drawn guess. The chain was then optimized by minimizing an appropriately designed cost function, composed of three contributions. The first term rewards a maximum contrast between the wall and the background. The second contribution compensates for the different background levels within the two adjacent domains. The third term favors straight domain walls and effectively sets a lower limit for the curvature radius. The resulting curve was used to calculate the domain wall inclination angle $\alpha$ and the surface distance $d$. For the domain wall shape reconstruction process presented in Figure 4, the domain wall shape was expanded in a fourth order Taylor series $d(x) = \sum_{k=0}^{4} a_k x^k$. The coefficients $a_k$, as well as the intensity change prefactor $A$ and the background intensity $I_0$, were optimized with a basin-hopping algorithm to yield a similar SEM intensity as in the experiment as:

$$I(x) = I_0 + \frac{A}{d^2(x)} \underbrace{\frac{d'(x)}{\sqrt{1 + [d'(x)^2]}}}_{=\cos\alpha(x)} . \tag{4}$$

The basin approach is required to avoid trapping in local minima due to the noisy experimental data. In general, fitting $a_k$ and $A$ simultaneously overparameterizes the problem. Under ideal experimental conditions this can, in principle, be bypassed by determining the parameter $A$ from calibration. Instead, in this study, one point of the domain wall $(x_f, d_f)$ was determined from the cross-sectional data (Figure 3d) and used as an additional constraint on the $a_k$ coefficients.

**References**

1. Meier, D., Seidel, J., Gregg, M., Ramesh, R.: Domain Walls: From Fundamental Properties to Nanotechnology Concepts. Oxford University Press (2020). https://doi.org/10.1093/oso/9780198862499.001.0001

2. Nataf, G.F., Guennou, M., Gregg, J.M., Meier, D., Hlinka, J., Salje, E.K.H., Kreisel, J.: Domain-wall engineering and topological defects in ferroelectric and ferroelastic materials. Nature Reviews Physics. 2, 634–648 (2020). https://doi.org/10.1038/s42254-020-0235-z

3. Bednyakov, P.S., Sturman, B.I., Sluka, T., Tagantsev, A.K., Yudin, P. V.: Physics and applications of charged domain walls. NPJ Comput Mater. 4, (2018). https://doi.org/10.1038/s41524-018-0121-8

4. Mundy, J.A., Schaab, J., Kumagai, Y., Cano, A., Stengel, M., Krug, I.P., Gottlob, D.M., Doğanay, H., Holtz, M.E., Held, R., Yan, Z., Bourret, E., Schneider, C.M., Schlom, D.G., Muller, D.A., Ramesh, R., Spaldin, N.A., Meier, D.: Functional electronic inversion layers at ferroelectric domain walls. Nat Mater. 16, 622–627 (2017). https://doi.org/10.1038/nmat4878

5. Sluka, T., Tagantsev, A.K., Bednyakov, P., Setter, N.: Free-electron gas at charged domain walls in insulating BaTiO3. Nat Commun. 4, (2013). https://doi.org/10.1038/ncomms2839

6. Salje, E.K.H.: Multiferroic domain boundaries as active memory devices: Trajectories towards domain boundary engineering. ChemPhysChem. 11, 940–950 (2010). https://doi.org/10.1002/cphc.200900943

7. Catalan, G., Seidel, J., Ramesh, R., Scott, J.F.: Domain wall nanoelectronics. Rev Mod Phys. 84, 119–156 (2012). https://doi.org/10.1103/RevModPhys.84.119

8. Meier, D., Selbach, S.M.: Ferroelectric domain walls for nanotechnology. Nat Rev Mater. 7, 157–173 (2022). https://doi.org/10.1038/s41578-021-00375-z

9. Sharma, P., Moise, T.S., Colombo, L., Seidel, J.: Roadmap for Ferroelectric Domain Wall Nanoelectronics. Adv Funct Mater. 32, (2022). https://doi.org/10.1002/adfm.202110263

10. Seidel, J., Martin, L.W., He, Q., Zhan, Q., Chu, Y.-H.H., Rother, A., Hawkridge, M.E., Maksymovych, P., Yu, P., Gajek, M., Balke, N., Kalinin, S. V., Gemming, S., Wang, F., Catalan, G., Scott, J.F., Spaldin, N.A., Orenstein, J., Ramesh, R.: Conduction at domain walls in oxide multiferroics. Nat Mater. 8, 229–234 (2009). https://doi.org/10.1038/nmat2373

11. Schaab, J., Skjærvø, S.H., Krohns, S., Dai, X., Holtz, M.E., Cano, A., Lilienblum, M., Yan, Z., Bourret, E., Muller, D.A., Fiebig, M., Selbach, S.M., Meier, D.: Electrical half-wave rectification at ferroelectric domain walls. Nat Nanotechnol. 13, 1028–1034 (2018). https://doi.org/10.1038/s41565-018-0253-5

12. Vul, B.M., Guro, G.M., Ivanchik, I.I.: Encountering domains in ferroelectrics. Ferroelectrics. 6, 29–31 (1973). https://doi.org/10.1080/00150197308237691

13. Eliseev, E.A., Morozovska, A.N., Svechnikov, G.S., Gopalan, V., Shur, V.Y.: Static conductivity of charged domain walls in uniaxial ferroelectric semiconductors. Phys Rev B. 83, 1–8 (2011). https://doi.org/10.1103/PhysRevB.83.235313

14. Gureev, M.Y., Tagantsev, A.K., Setter, N.: Head-to-head and tail-to-tail 180° domain walls in an isolated ferroelectric. Phys Rev B Condens Matter Mater Phys. 83, 1–18 (2011). https://doi.org/10.1103/PhysRevB.83.184104

15. Meier, D., Seidel, J., Cano, A., Delaney, K., Kumagai, Y., Mostovoy, M., Spaldin, N.A., Ramesh, R., Fiebig, M.: Anisotropic conductance at improper ferroelectric domain walls. Nat Mater. 11, 284–288 (2012). https://doi.org/10.1038/nmat3249

16. Kämpfe, T., Reichenbach, P., Schröder, M., Haußmann, A., Eng, L.M., Woike, T., Soergel, E.: Optical three-dimensional profiling of charged domain walls in ferroelectrics by Cherenkov second-harmonic generation. Phys Rev B Condens Matter Mater Phys. 89, 035314 (2014). https://doi.org/10.1103/PhysRevB.89.035314

17. Sheng, Y., Best, A., Butt, H.J., Krolikowski, W., Arie, A., K. Koynov: Three-dimensional ferroelectric domain visualization by Cerenkov-type second harmonic generation. Opt.Express. 18, 16539–16545 (2010). https://doi.org/10.1364/OE.18.016539

18. Cherifi-Hertel, S., Bulou, H., Hertel, R., Taupier, G., Dorkenoo, K.D.H., Andreas, C., Guyonnet, J., Gaponenko, I., Gallo, K., Paruch, P.: Non-Ising and chiral ferroelectric domain walls revealed by nonlinear optical microscopy. Nat Commun. 8, 15768 (2017). https://doi.org/10.1038/ncomms15768

19. Acevedo-Salas, U., Croes, B., Zhang, Y., Cregut, O., Kokou Dodzi Dorkenoo, Kirbus, B., Singh, E., Beccard, H., Rüsing, M., Eng, L.M., Hertel, R., Eliseev, E.A., Morozovska, A.N., Cherifi-Herte, S.: Impact of 3D Curvature on the Polarization Orientation in Non-Ising Domain Walls. Nano Lett. 23, 795–803 (2023). https://doi.org/10.1021/acs.nanolett.2c03579

20. Jia, C.L., Mi, S.B., Urban, K., Vrejoiu, I., Alexe, M., Hesse, D.: Atomic-scale study of electric dipoles near charged and uncharged domain walls in ferroelectric films. Nat Mater. 7, 57–61 (2008). https://doi.org/10.1038/nmat2080

21. Småbråten, D.R., Holstad, T.S., Evans, D.M., Yan, Z., Bourret, E., Meier, D., Selbach, S.M.: Domain wall mobility and roughening in doped ferroelectric hexagonal manganites. Phys Rev Res. 2, 033159 (2020). https://doi.org/10.1103/PhysRevResearch.2.033159

22. Paruch, P., Giamarchi, T., Triscone, J.M.: Domain wall roughness in epitaxial ferroelectric $PbZr_{0.2}Ti_{0.8}O_3$ thin films. Phys Rev Lett. 94, 197601 (2005). https://doi.org/10.1103/PhysRevLett.94.197601

23. Roede, E.D., Shapovalov, K., Moran, T.J., Mosberg, A.B., Yan, Z., Bourret, E., Cano, A., Huey, B.D., van Helvoort, A.T.J., Meier, D.: The Third Dimension of Ferroelectric Domain Walls. Advanced Materials. 34, 1–13 (2022). https://doi.org/10.1002/adma.202202614

24. Song, J., Zhou, Y., Huey, B.D.: 3D structure-property correlations of electronic and energy materials by tomographic atomic force microscopy. Appl Phys Lett. 118, 050801 (2021). https://doi.org/10.1063/5.0040984

25. Song, J., Zhuang, S., Martin, M., Ortiz-Flores, L.A., Paudel, B., Yarotski, D., Hu, J., Chen, A., Huey, B.D.: Interfacial-Strain-Controlled Ferroelectricity in Self-Assembled $BiFeO_3$ Nanostructures. Adv Funct Mater. 31, 2102311 (2021). https://doi.org/10.1002/adfm.202102311

26. McCluskey, C.J., Kumar, A., Gruverman, A., Luk'yanchuk, I., Gregg, J.M.: Domain wall saddle point morphology in ferroelectric triglycine sulfate. Appl Phys Lett. 122, 222902 (2023). https://doi.org/10.1063/5.0152518

27. Smolenskii, G.A., Chupis, I.E.: Ferroelectromagnets. Soviet Physics Uspekhi. 25, 475–493 (1982). https://doi.org/10.1070/PU1982v025n07ABEH004570

28. Hunnestad, K.A., Roede, E.D., Helvoort, A.T.J. van, Meier, D.: Characterization of ferroelectric domain walls by scanning electron microscopy. J Appl Phys. 128, 191102 (2020). https://doi.org/10.1063/5.0029284

29. Schultheiß, J., Schaab, J., Småbråten, D.R., Skjærvø, S.H., Bourret, E., Yan, Z., Selbach, S.M., Meier, D.: Intrinsic and extrinsic conduction contributions at nominally neutral domain walls in hexagonal manganites. Appl Phys Lett. 116, 262903 (2020). https://doi.org/10.1063/5.0009185

30. Schoenherr, P., Shapovalov, K., Schaab, J., Yan, Z., Bourret, E.D., Hentschel, M., Stengel, M., Fiebig, M., Cano, A., Meier, D.: Observation of Uncompensated Bound Charges at Improper Ferroelectric Domain Walls. Nano Lett. 19, 1659–1664 (2019). https://doi.org/10.1021/acs.nanolett.8b04608

31. Choi, T., Horibe, Y., Yi, H.T., Choi, Y.J., Wu, W., Cheong, S.W.: Insulating interlocked ferroelectric and structural antiphase domain walls in multiferroic $YMnO_3$. Nat Mater. 9, 253–258 (2010). https://doi.org/10.1038/nmat2632

32. Wu, W., Horibe, Y., Lee, N., Cheong, S.W., Guest, J.R.: Conduction of topologically protected charged ferroelectric domain walls. Phys Rev Lett. 108, 077203 (2012). https://doi.org/10.1103/PhysRevLett.108.077203

33. Kumagai, Y., Spaldin, N.A.: Structural domain walls in polar hexagonal manganites. Nat Commun. 4, 1–8 (2013). https://doi.org/10.1038/ncomms2545

34. Campbell, M.P., McConville, J.P.V., McQuaid, R.G.P., Prabhakaran, D., Kumar, A., Gregg, J.M.: Hall effect in charged conducting ferroelectric domain walls. Nat Commun. 7, 13764 (2016). https://doi.org/10.1038/ncomms13764

35. Maksymovych, P., Morozovska, A.N., Yu, P., Eliseev, E.A., Chu, Y.H., Ramesh, R., Baddorf, A.P., Kalinin, S. V.: Tunable metallic conductance in ferroelectric nanodomains. Nano Lett. 12, 209–213 (2012). https://doi.org/10.1021/nl203349b

36. McConville, J.P.V., Lu, H., Wang, B., Tan, Y., Cochard, C., Conroy, M., Moore, K., Harvey, A., Bangert, U., Chen, L.Q., Gruverman, A., Gregg, J.M.: Ferroelectric Domain Wall Memristor. Adv Funct Mater. 30, 2000109 (2020). https://doi.org/10.1002/adfm.202000109

37. Robinson, G.Y., White, R.M.: Scanning electron microscopy of ferroelectric domains in barium titanate. Appl Phys Lett. 10, 320–323 (1967). https://doi.org/10.1063/1.1754829

38. Meyer, K.P., Blumtritt, H., Szczesniak, L.: Visualization of domain boundaries in $Gd_2(MoO_4)_3$ single crystals by scanning electron microscope potential contrast. Ultramicroscopy. 6, 67–70 (1981)

39. Aristov, V. V., Kokhanchik, L.S., Voronovskii, Y.I.: Voltage contrast of ferroelectric domains of Lithium Niobate in SEM. Physica Status Solidi (a). 86, 133–141 (1984). https://doi.org/10.1002/pssa.2210860113

40. Li, J.Q., Yang, H.X., Tian, H.F., Ma, C., Zhang, S., Zhao, Y.G., Li, J.Q.: Scanning secondary-electron microscopy on ferroelectric domains and domain walls in YMnO$_3$. Appl Phys Lett. 100, 152903 (2012). https://doi.org/10.1063/1.4704165

41. Roede, E.D., Mosberg, A.B., Evans, D.M., Bourret, E., Yan, Z., Van Helvoort, A.T.J.J., Meier, D.: Contact-free reversible switching of improper ferroelectric domains by electron and ion irradiation. APL Mater. 9, 021105 (2021). https://doi.org/10.1063/5.0038909

42. Zou, Y.B., Mao, S.F., Da, B., Ding, Z.J.: Surface sensitivity of secondary electrons emitted from amorphous solids: Calculation of mean escape depth by a Monte Carlo method. J Appl Phys. 120, (2016). https://doi.org/10.1063/1.4972196

43. Goldstein, J.I., Newbury, D.E., Michael, J.R., Ritchie, N.W.M., Scott, J.H.J., Joy, D.C.: Scanning electron microscopy and X-Ray Microanalysis. Springer Nature (2018). https://doi.org/10.1007/978-1-4939-6676-9

44. Zhao, M., Ming, B., Kim, J.W., Gibbons, L.J., Gu, X., Nguyen, T., Park, C., Lillehei, P.T., Villarrubia, J.S., Vladár, A.E., Alexander Liddle, J.: New insights into subsurface imaging of carbon nanotubes in polymer composites via scanning electron microscopy. Nanotechnology. 26, (2015). https://doi.org/10.1088/0957-4484/26/8/085703

45. Yan, Z., Meier, D., Schaab, J., Ramesh, R., Samulon, E., Bourret, E.: Growth of high-quality hexagonal ErMnO$_3$ single crystals by the pressurized floating-zone method. J Cryst Growth. 409, 75–79 (2015). https://doi.org/10.1016/j.jcrysgro.2014.10.006

46. Mosberg, A.B., Roede, E.D., Evans, D.M., Holstad, T.S., Bourret, E., Yan, Z., Van Helvoort, A.T.J., Meier, D.: FIB lift-out of conducting ferroelectric domain walls in hexagonal manganites. Appl Phys Lett. 115, 122901 (2019). https://doi.org/10.1063/1.5115465

47. Jungk, T., Hoffmann, Á., Fiebig, M., Soergel, E.: Electrostatic topology of ferroelectric domains in YMnO$_3$. Appl Phys Lett. 97, 012904 (2010). https://doi.org/10.1063/1.3460286

48. Wu, W., Guest, J.R., Horibe, Y., Park, S., Choi, T., Cheong, S.W., Bode, M.: Polarization-modulated rectification at ferroelectric surfaces. Phys Rev Lett. 104, 217601 (2010). https://doi.org/10.1103/PhysRevLett.104.217601

49. Bihan, R. Le: Study of ferroelectric and ferroelastic domain structures by scanning electron microscopy. Ferroelectrics. 97, 19–46 (1989). https://doi.org/10.1080/00150198908018081

50. Srinivasan, A., Han, W., Khursheed, A.: Secondary electron energy contrast of localized buried charge in metal-insulator-silicon structures. Microscopy and Microanalysis. 24, 453–460 (2018). https://doi.org/10.1017/S1431927618015052

51. Drouin, D., Couture, A.R., Joly, D., Tastet, X., Aimez, V., Gauvin, R.: CASINO V2.42 - A fast and easy-to-use modeling tool for scanning electron microscopy and microanalysis users. Scanning. 29, 92–101 (2007). https://doi.org/10.1002/sca.20000

**Acknowledgements**

The authors thank J. Schultheiß for fruitful discussions. D.M. thanks NTNU for support through the Onsager Fellowship Program and the Outstanding Academic Fellow Program. D.M., J.H., L.R., and U.L. acknowledge funding from the European Research Council (ERC) under the European Union's Horizon 2020 Research and Innovation Program (Grant Agreement No. 863691). M.Z. acknowledges funding from the Studienstiftung des Deutschen Volkes via a doctoral grant and the State of Bavaria via a Marianne-Plehn scholarship. G.C. acknowledges financial support from grant number PRX22/00595, from the Spanish Ministry of Universities, during part of his stay at NTNU in Trondheim. The Research Council of Norway is acknowledged for the support to the Norwegian Micro- and Nano-Fabrication Facility, Nor-Fab, project number 295864.

**Contributions**

J.H. performed the scanning probe microscopy (SPM), scanning electron microscopy (SEM), and focused ion beam (FIB) experiments, analyzed the data, and proposed the model describing the SEM contrast, supervised by D.M. M.Z. refined the model and developed the theoretical framework for the 3D reconstruction, supervised by I.K. and D.M., with support from I.U.; L.R., U.L., E.D.R. and G.C. contributed to the interpretation of the data. Z.Y. and E.B. provided the materials. D.M. devised and coordinated the project. J.H. and D.M. wrote the manuscript, supported by M.Z. concerning the 3D reconstruction. All the authors discussed the results and contributed to the final version of the manuscript.

**Corresponding author**

Dennis Meier, dennis.meier@ntnu.no

# Supplementary Information

# Non-destructive tomographic nanoscale imaging of ferroelectric domain walls

Jiali He[1], Manuel Zahn[1,2], Ivan N. Ushakov[1], Leonie Richarz[1], Ursula Ludacka[1], Erik D. Roede[1], Zewu Yan[3,4], Edith Bourret[4], István Kézsmárki[2], Gustau Catalan[1,5,6], Dennis Meier[1]

[1]Department of Materials Science and Engineering, Norwegian University of Science and Technology (NTNU), Trondheim, Norway.

[2]Experimental Physics V, Center for Electronic Correlation and Magnetism, Universität Augsburg, Augsburg, Germany.

[3]Department of Physics, ETH Zurich, Zurich, Switzerland.

[4]Materials Sciences Division, Lawrence Berkeley National Laboratory, Berkeley, USA.

[5]Catalan Institute of Nanoscience and Nanotechnology (ICN2), Campus Universitat Autonoma de Barcelona, Bellaterra, Catalonia.

[6]ICREA-Institucio Catalana de Recerca i Estudis Avançats, Barcelona, Catalonia.

**Supplementary Note 1: Determination of exponent $n$**

To determine the exponent $n$ in Equation (2) of the main text and test if the ansatz describes the measured SEM intensity variations, we extract the parameters $\alpha$, $d$ and $I_{\text{exp}}$ for the domain walls DW1 and DW2 in Figure 3d. The change in intensity is $\Delta I_{\text{exp}} = I_{\text{exp}} - I_{\text{exo,bg}}$, where $I_{\text{exp,bg}}$ denotes the background intensity. Equation (3) can be brought into the following form, leading to a linear relation with respect to $n$:

$$\ln \Delta I_{\text{exp}} - \ln \cos \alpha = \ln A - n \cdot \ln d. \qquad (S1)$$

The obtained representation is visualized for DW1 and DW2 in Supplementary Figure S1. In both cases, the data is reasonably well described by the ansatz, demonstrating its general validity. Based on the slope, we find n = 2.31 ± 0.09 for DW1 and n = 2.35 ± 0.08 for DW2. We thus conclude that $n = 2$ is a good approximation, while the numerical uncertainties are underestimated. For the intensity amplitudes, we find $A = 0.726 \pm 0.019$ for DW1 and $A = 1.169 \pm 0.030$ for DW2. The resulting intensities $I_{\text{calc}}$ (black), calculated based on Equation (3) in the main text ($I_{\text{calc}} \equiv \Delta I$), are plotted together with the experimental SEM intensity $I_{\text{exp}}$ data in Supplementary Figure S2 (tail-to-tail: $\cos \alpha > 0$, head-to-head: $\cos \alpha < 0$); values for $\alpha$ and $d$ are extracted from the data in Figure 3d. The experimental SEM intensity data, obtained between points A and B in Figure 3c (red dotted line, 50 pixels) was smoothed using the Savitzky-Golay method with a window size of 40 and polynomial order of 2. Subsequently, the subsequent was normalized ($x_{\text{normalized}} = \frac{x - x_{\text{min}}}{x_{\text{max}} - x_{\text{min}}}$; x represents individual data points, $x_{\text{min}}$ and $x_{\text{max}}$ denote the dataset's minimum and maximum value).

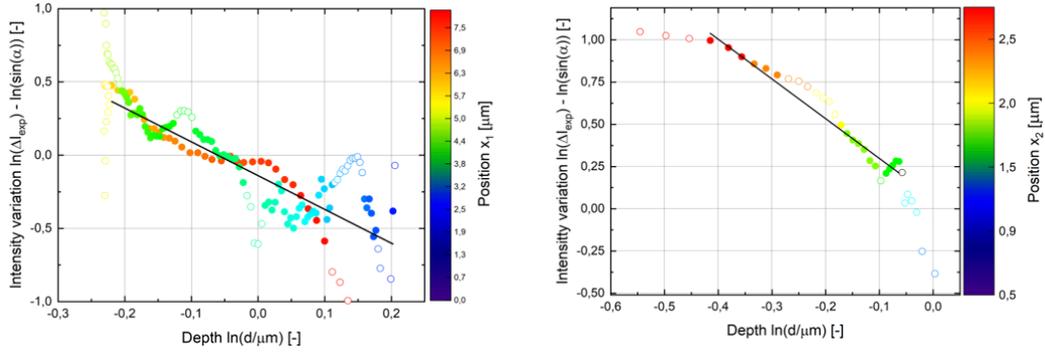

**Figure S1 | Determination of exponent *n* based on two datasets for the quantitative model.** Based on the cross section shown in Figure 3d in the main text, the different parameters describing the domain-wall shape are determined for two domain walls, i.e., DW1 (left panel) and DW2 (right panel). Datapoints that have been excluded for the linear fitting process as they aren't part of the linear regime are visualized with open circles.

**Supplementary Note 2: Direct calculation of surface intensity**

In contrast to the fitting approach described in the manuscript, the agreement of the experimental data with the (specified) model developed in Equation (3) can be verified in a more direct way. To do so, the expression $\cos \alpha / d^2$ is directly calculated from the cross-section data using numerical differentiation and the explicit version of the model developed in the methods section of the manuscript. By adjusting $A$ and $I_0$ manually, the qualitative agreement of experimental and directly calculated SEM intensity can be tested and is shown for DW1 and DW2 in figure S2. This approach confirms the result of the main text that the specified model well-describes the experimental findings.

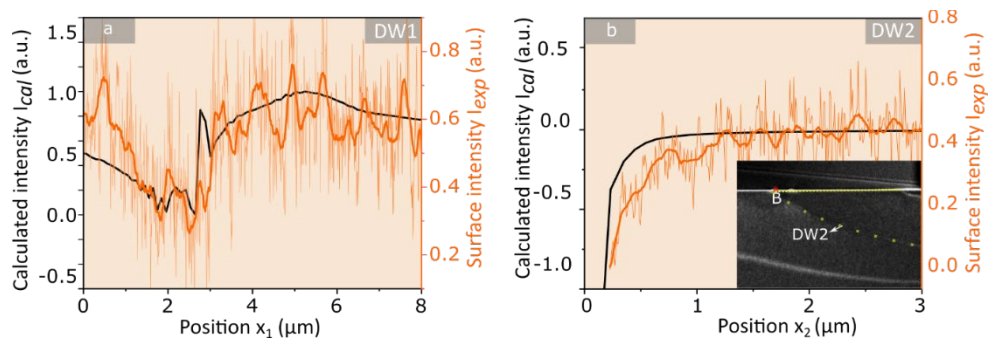

**Figure S2 | Comparison of measured and calculated SEM surface intensity. a,** Comparison of the SEM intensity $I_{\text{exp}}$ measured along the red dashed line between A and B in Figure 3c in the main text and the calculated intensity $I_{\text{calc}}$ for DW1 based on direct numerical calculation. **b,** Same as in **a** for the yellow dashed line in Figure 3 and DW2 (see also inset to **b**).